\documentclass[twocolumn,conference]{IEEEtran}
\usepackage{amsmath}
\usepackage{graphicx}
\usepackage{subcaption}
\usepackage{cite}
\usepackage{epstopdf}
\usepackage{balance}
\usepackage{gensymb}
\usepackage{color}
\usepackage{lipsum}
\usepackage{float}
\usepackage{epstopdf}
\usepackage[mathcal]{eucal}
\usepackage{subfiles}
\usepackage[T1]{fontenc}
\usepackage{tikz}
\usepackage{pgfplots}

\newcommand{\ignore}[1]{}

\IEEEoverridecommandlockouts

\begin{document}

\bstctlcite{IEEEexample:BSTcontrol}

\title{Understanding 
End-to-End Effects of Channel Dynamics 
in Millimeter Wave 5G New Radio}

\author{Christopher Slezak, Menglei Zhang, Marco Mezzavilla, and Sundeep Rangan
\\\IEEEauthorblockA{\{chris.slezak, menglei, mezzavilla.marco, srangan\}@nyu.edu} \thanks{This work is supported in part by the National Science Foundation under Grants 1302336, 1564142, and 1547332, NIST grant 70NANB17H166, the Semiconductor Research Corporation and NYU Wireless.}}

\maketitle

\begin{abstract}
A critical challenge for wireless communications
in the millimeter wave (mmWave) bands is blockage.  MmWave signals
suffer significant penetration losses from many common materials and objects, and small changes in the position of obstacles in the environment can cause large variations in the channel quality.
This paper provides a measurement-based
study of the effects of human blockage on an end-to-end
application over a mmWave 
cellular link.  A phased array system is used to measure the channel in multiple directions almost simultaneously
in a realistic indoor scenario.  
The measurements
are integrated into a detailed ns-3 simulation that models
both the latest 3GPP New Radio beam search procedure as well
as the internet protocol stack. The 
measurement-based simulation 
illustrates how recovery from blockage depends
on the path diversity and beam search.
\end{abstract}


\section{Introduction}

The millimeter wave (mmWave) bands and other frequencies above 6~GHz
have become a key component of the emerging fifth generation (5G) cellular standards
\cite{3GPP38.300,rappaportmillimeter,RanRapE:14}.
These frequencies offer vastly greater bandwidths than what is available in the conventional 
sub-6~GHz bands providing the potential to meet the massive mobile broadband 
and ultra-low latency requirements of the 5G vision \cite{dahlman20145g}.
However, a key challenge for the use of the mmWave bands
in mobile applications is blockage:  
MmWave signals are significantly more vulnerable than signals at conventional frequencies
to blockage by many common obstacles and materials
\cite{Rappaport2014-mmwbook}.

In addition to the reduced coverage and range, the susceptibility to blockage of mmWave signals can 
result in highly variable link quality.  Small motion of the handset relative to
blockers and obstacles in the environment (including the hand or body)
can result in dramatic swings in received power.
The 
effects of blockage are complicated by the directional
nature of the transmission in the mmWave 
frequencies.  In cellular applications, 
mmWave signals will be transmitted and received in highly directional beams
to overcome the high isotropic path loss in these frequencies.  Blockage may
be mitigated by finding alternate unblocked directions of communication.  However,
the performance of these techniques depends on the spatial diversity of the channel,
the latency in beam tracking protocols and signal processing algorithms as well as the application
and transport-layer 
effects of disruptions in link quality while alternate paths are being discovered.  

The broad purpose of this paper is to provide an \emph{end-to-end perspective} on 
the effects of blockage in 5G mmWave cellular systems.
In particular, we evaluate the interaction of three underlying processes operating
at different levels in the protocol stack:
\begin{itemize}
  \item \emph{Experimentally measured channel dynamics}:  
  Building on our measurement system in \cite{slezak60GHzBlockage},
  we experimentally measure blockage in a complex, but realistic, blockage scenario
  in a home entertainment-type setting.  Importantly, our measurement system can 
  measure the channel in multiple directions nearly simultaneously, thereby providing a complete
  spatial and temporal trace of the channel.
  
  \item \emph{3GPP NR beam search:} The channel measurements
  are then incorporated into a detailed ns-3 based simulation 
  \cite{mezzavilla20155g,ford2016framework}
  along with the latest 3GPP New Radio (NR) standard
  \cite{3GPP38.300}.  The simulation accurately 
  models the beam tracking procedure at both the base station (gNB in NR terminology) and mobile user equipment (UE).  
  The simulation considers both 
  analog and digital beamforming.
  
  \item \emph{Internet protocol stack:}  The simulation also includes a detailed and realistic model
  of the core network as well as widely-used TCP congestion control algorithms. 
  
\end{itemize}

The study thus provides the most realistic modeling
of blockage in an end-to-end evaluation mmWave cellular system to date.
Most significantly,  we include spatial dynamic measurements 
in a complex and realistic blocking scenario.  
While there are a large number of works precisely measuring penetration losses of mmWave signals
with various materials and the human body \cite{gustafson2011HumanShad60G,maccartney73Gblockage,weiler2016EnvironmentInducedShadowing},
the time dynamics of blockage are less understood. 
Most prior works studying blockage dynamics have measurement systems with omnidirectional antennas that provide no spatial information, fixed horn antennas that measure
the blockage in a single direction, or MIMO measurements with a small number of antennas \cite{giordani2016channel,jacob11ad60GModel,maccartneyRapidFadingCrowds,Collonge04,peter11blockmeas}.  However, to fully understand
the path diversity and beam tracking, it is necessary to capture the channel dynamics
over multiple paths.  For this \textit{}purpose, we use a phased array system developed in our
earlier work \cite{slezak60GHzBlockage} that measures the channel in multiple directions.
While \cite{slezak60GHzBlockage} studied the channel in a laboratory environment,
this study is based on a more extensive measurement campaign with a realistic indoor scenario and multiple
moving blockers. Phased array systems have also been used for hand blockage in a recent
Qualcomm study \cite{qualcomm5GNRBlockage}.

The blockage traces we measure experimentally are then integrated into a detailed
ns-3 simulation developed in our earlier work \cite{mezzavilla20155g,ford2016framework}.  
The simulator accurately models all layers of the communication stack 
including MAC, RLC, and PDCP as well as the core network
and transport layer protocols and can capture the effects of rate adaptation, hybrid ARQ
and RLC retransmissions during the blockage events.  
This simulator has been used in several prior works on transport
layer \cite{zhangTransportPerformance,ford2017achieving,zhang2017tcp},
but used a hypothetical frame structure \cite{dutta2017frame}.  In this work, we integrate
the 3GPP NR beam search procedures as described in \cite{3GPP38.300}.

One of the interesting features of the latest 3GPP NR specification is the relatively infrequent
transmission of the synchronization signal (SS) bursts for channel tracking.
The infrequent transmission of
SS signals motivates consideration of 
digital beamforming, which
can track multiple directions simultaneously.  
Prior work \cite{barati2015directional,barati2016initial}
has shown that digital beamforming can provide
much faster initial access.  However, we will see
that the gains for blockage may be less.

\section{Measurement System}
To investigate the effects of mmWave blockage, 
our earlier work \cite{slezak60GHzBlockage}
developed a measurement system using two SiBeam 60 GHz phased antenna arrays.
The arrays consist of 12 radiating elements with the ability to adjust the phase on each element with 2 bits of precision. Modifying these phases alters the radiation pattern of the array and allows the user to "steer" the antenna array without any moving parts such as a gimbal. A new beamforming vector can be applied in less than one microsecond, which is in sharp contrast to mechanical methods of steering which are typically on the order of tens of milliseconds\cite{FLIRmanual}. The ability to rapidly switch between different steering vectors enables measurements to be conducted over multiple directions on the same timescale as dynamic events such as human blockage.

\subsection{Hardware Components}
\begin{figure}
\includegraphics[width=\linewidth]{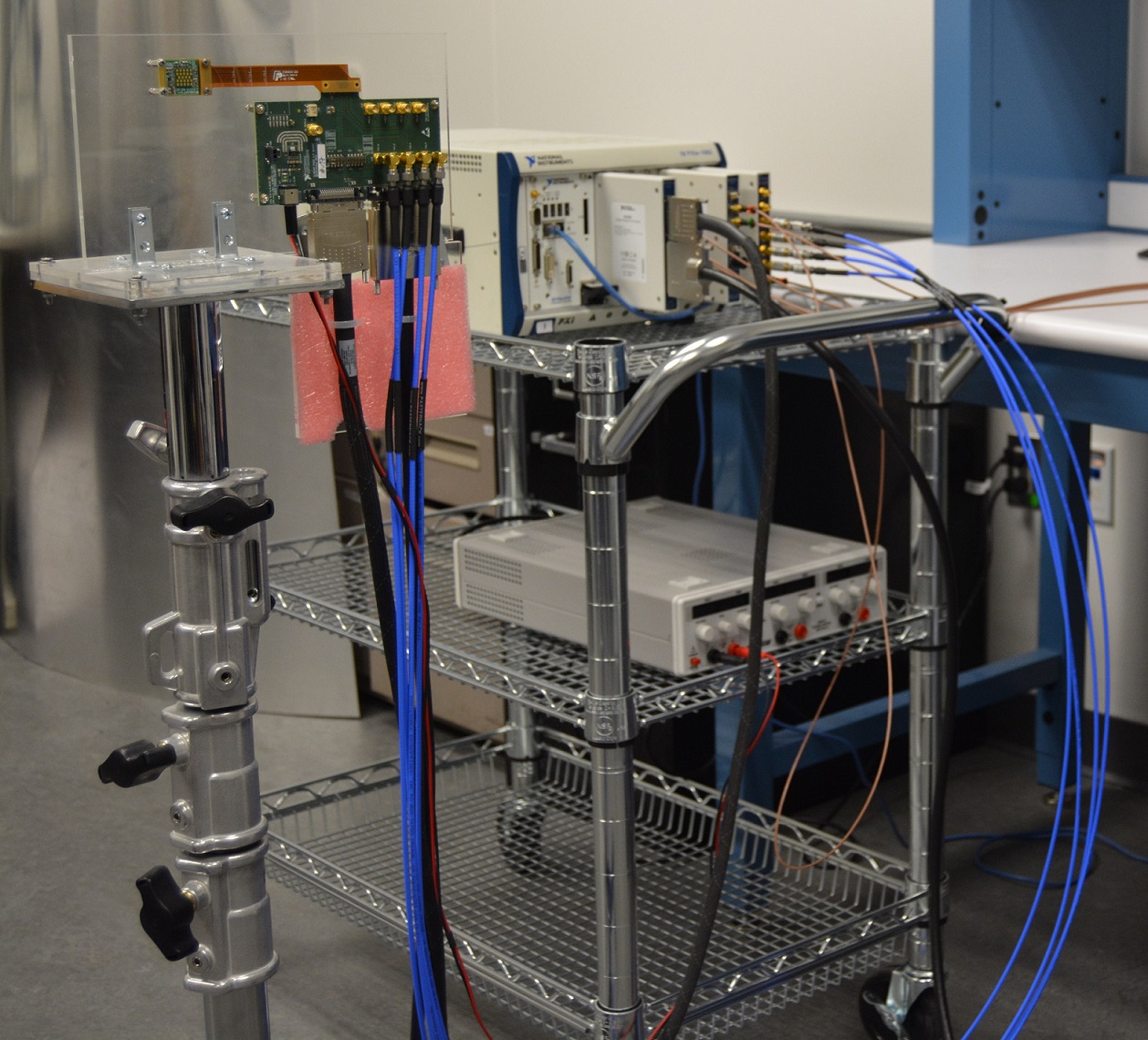}
\caption{Transmitter portion of the measurement system showing the PXIe chassis and SiBeam array.}
\label{fig:TX}
\end{figure}

To support the SiBeam phased arrays, the measurement system includes two National Instruments (NI) PCI eXtensions for Instumentation Express (PXIe) chassis which perform all baseband processing and necessary communication to maintain synchronization during a measurement. The two chassis are each equipped with several Field Programmable Gate Array (FPGA) modules and Input/Output (I/O) daughterboards that allow the FPGA modules to send baseband and control signals to the arrays as well as share timing signals between the two chassis. A chassis controller running a real-time operating system is also present to send commands to the FPGA modules and to allow the two chassis to communicate via Ethernet. The transmitter (TX) portion of the measurement system is shown in Fig. \ref{fig:TX}.

\subsection{Measurement Procedure}
Steering the array is accomplished with a codebook of twelve vectors that define the phases of the elements in the array. To begin, the TX will apply one steering vector, and the RX will cycle through its entire codebook of twelve steering vectors. For each of these, the RX will acquire a single power delay profile (PDP) which describes the multipath characteristics of the wireless channel. The TX will then advance to the next steering vector and the RX repeats the same procedure to acquire twelve more PDPs, one for each of the steering vectors in its codebook. When the TX has finished sweeping over its whole codebook, we have in total 144 PDPs that were each acquired for a different combination of TX and RX steering vectors, or pointing angle combination (PAC).

This 144 PAC scan be performed in less than 1 millisecond due to the very rapid steering ability of the phased arrays. Critically, this is fast enough that a human walking at a typical speed will not move very far between scans, and it is therefore possible to observe the time evolution of attenuation due to human blockage across the multiple paths that are present in the channel. Due to memory constraints, the scans can only be repeated for a relatively short period of time. For the results shown in this paper, the scans were repeated over a period of 5.6 seconds, which is more than enough time for a single blockage event to begin and end.

\section{Blockage Experiments}
\begin{figure}
\includegraphics[width=\linewidth]{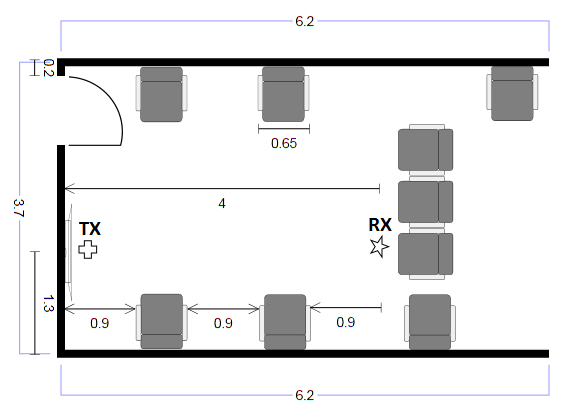}
\caption{Measurement environment as viewed from above. All dimensions are in meters.}
\label{fig:room}

\end{figure}

Measurements were performed in a room on the NYU Tandon campus chosen because it emulates a typical living room environment. The TX was placed below a wall-mounted television and the RX was located in front of a sofa 4 meters away from the TX. Both were at a height of 1 meter. Additional furniture in the room included several fabric chairs with attached tables. Fig. \ref{fig:room} shows the dimensions of the room as well as the locations of the TX, RX and furniture.

\begin{figure}
\begin{subfigure}{.49\linewidth}
\includegraphics[width=\linewidth]{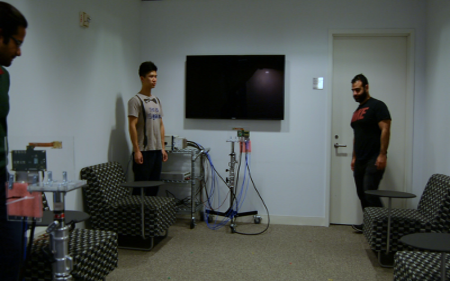}
\end{subfigure}
\begin{subfigure}{.49\linewidth}
\includegraphics[width=\linewidth]{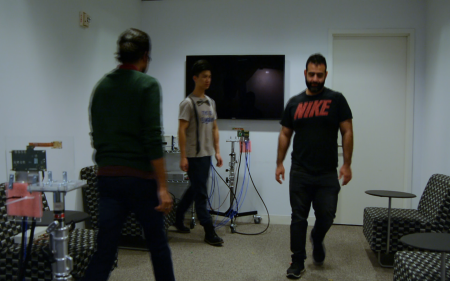}
\end{subfigure}

\vspace{3px}
\begin{subfigure}{.49\linewidth}
\includegraphics[width=\linewidth]{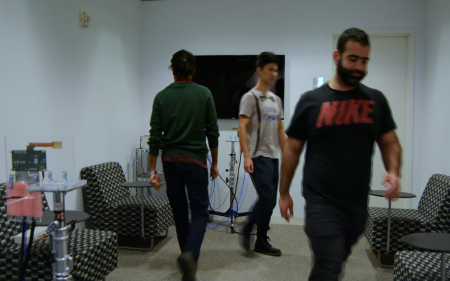}
\end{subfigure}
\begin{subfigure}{.49\linewidth}
\includegraphics[width=\linewidth]{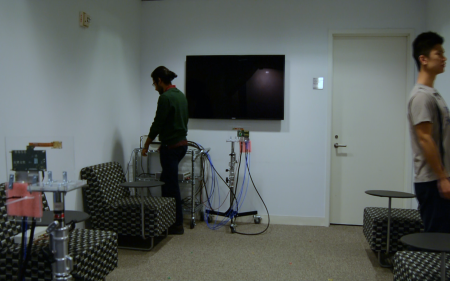}
\end{subfigure}

\caption{Screenshots from a video taken as blockers moved through the room during a measurement.}
\label{fig:screenshots}
\end{figure}

Measurements were performed for a variety of scenarios. The number of human blockers moving throughout the room varied from one to three, with the single blocker experiments performed with different distances between the blocker and the TX/RX. Some measurements were also performed with mobility at the RX in addition to blockage. For this work, we focus our analysis on a single trace from a measurement with three blockers walking through the room simultaneously. Fig. \ref{fig:screenshots} shows still images taken from a video of the blockers as they moved though the room.

\begin{figure}
\includegraphics[width=\linewidth]{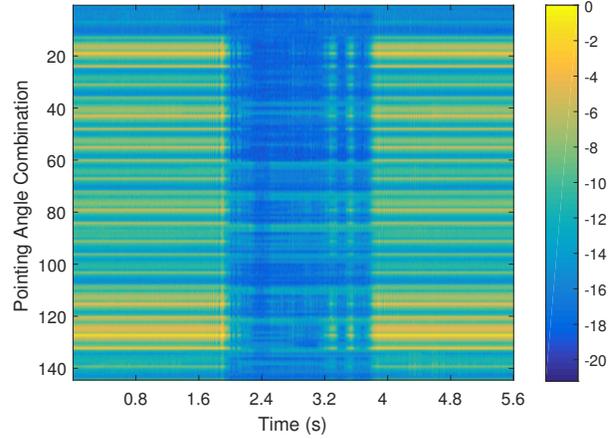}
\caption{Received power over time for each possible combination of TX and RX pointing angles.}
\label{fig:receivedPower}
\end{figure}

For this particular measurement, Fig. \ref{fig:receivedPower} shows the received power over time for all combination of TX and RX steering vectors. Received power is normalized to the strongest value observed when no paths are blocked. There is a very long blockage event visible in this figure, however note that some pointing angle combinations (those which direct some energy to a NLOS path in the channel) are not blocked at the same time as the more dominant LOS path. By incorporating this measurement trace into ns-3, we can explore how different beam tracking schemes are able to exploit this path diversity to overcome the effects of blockage.

\section{Modeling the 5G NR Beam Search}

\begin{figure}[!t]
\centering
\includegraphics[width=\columnwidth]{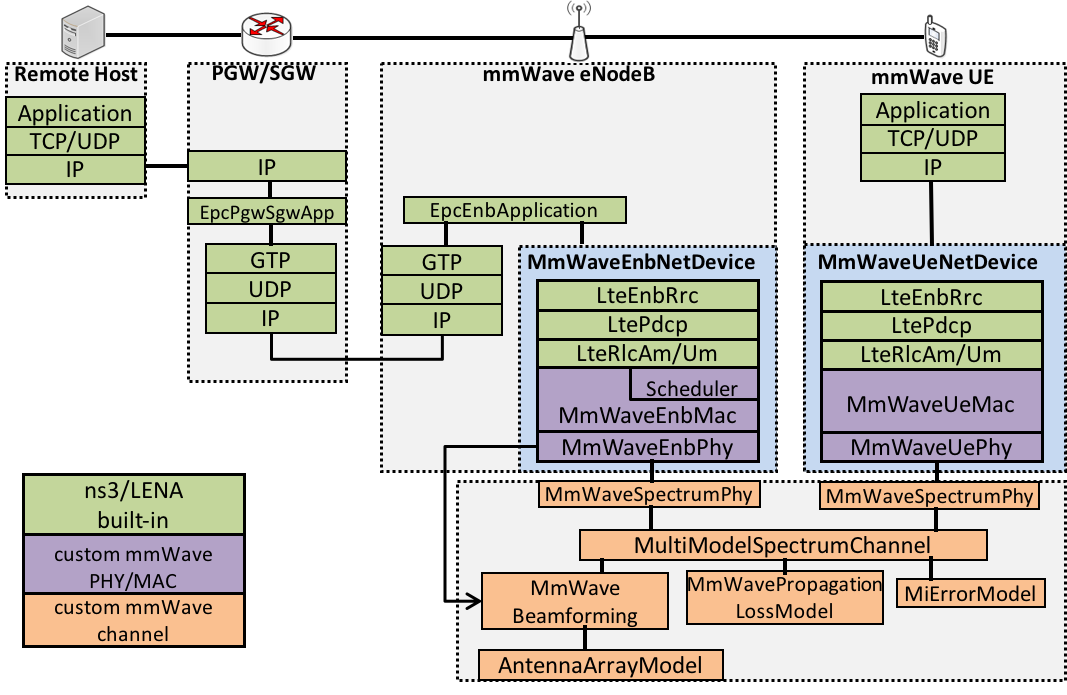}
\caption{Class diagram of the end-to-end mmWave module, as reported in \cite{ford2016framework}.}
\label{fig:framework}
\end{figure}

To understand the end-to-end effects of blockage,
the measured traces are incorporated into
the discrete-event network simulator ns-3 
with a mmWave module developed in our
earlier work \cite{mezzavilla20155g,ford2016framework}, as depicted in Fig. \ref{fig:framework}.
The simulator includes a complete model
for all PHY, MAC, RLC and PDCP layers in the RAN,
as well as the core networks protocols and delay.
In particular, both MAC and RLC retransmissions
during transmission losses are accounted for.

\begin{figure*}[ht!]
\centering
\includegraphics[width=0.95\textwidth]{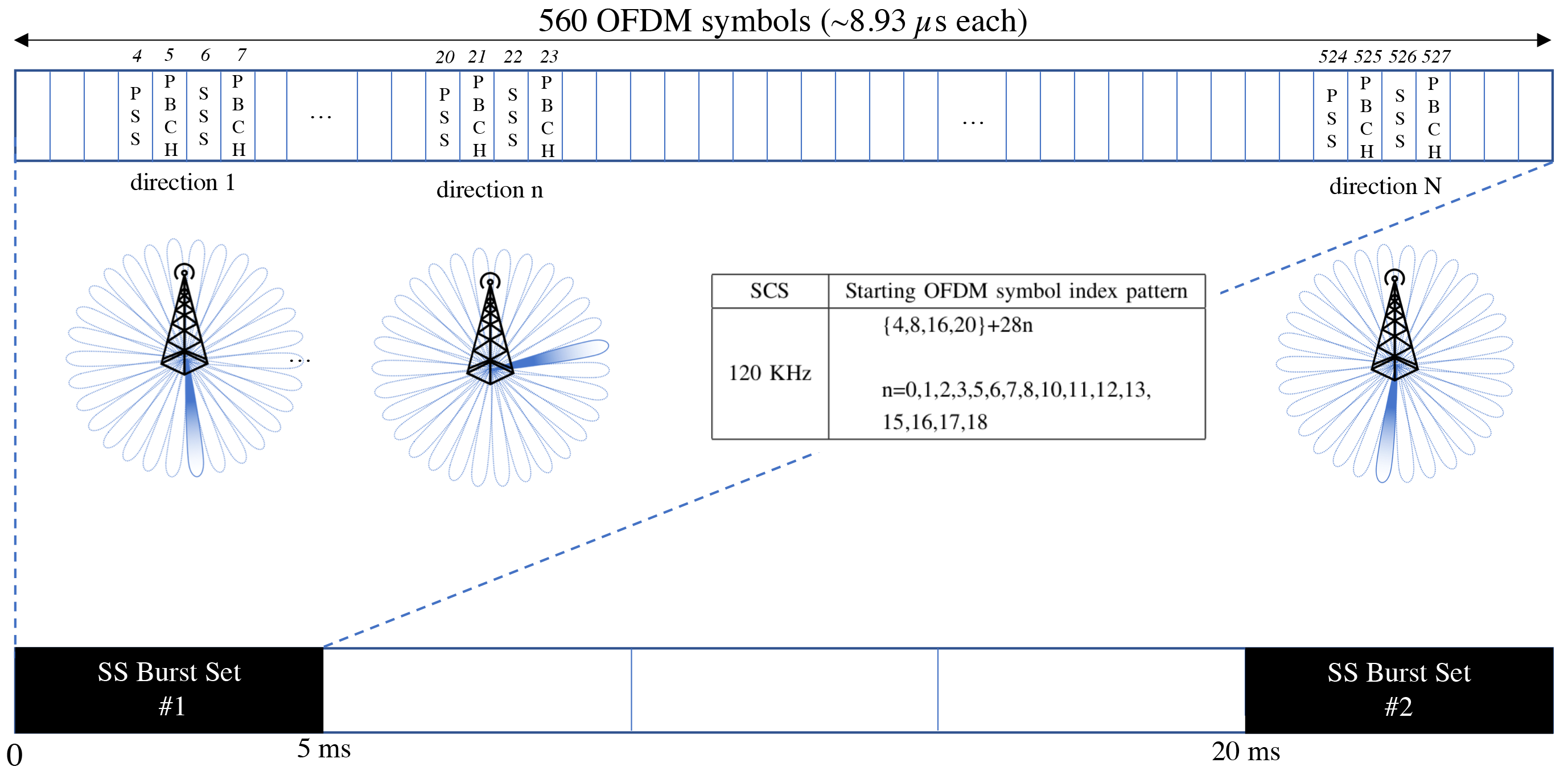}
\caption{SS Burst Set configuration.}
\label{fig:SSBs}
\end{figure*}

In this work, we have supplemented the model
with the NR beam management procedure as described in \cite{3GPP38.300} and shown in Fig. \ref{fig:SSBs}. In the NR standard, 
beam tracking is performed by the gNB cell periodically
transmitting a so-called synchronization signal
(SS) burst that scans across a
set of possible transmit directions.
In this study, we assume that both the gNB cell
and UE use phased array transceivers identical to those
used in the measurements.  Hence, each side 
must scan 12 directions.  At the gNB side,
we assume a default configuration where 
the SS burst set is repeated every 20 ms. 
Within one SS burst set, 
the gNB cell transmits pilots in all 12 directions over a specific set of OFDM symbols, as indicated in the table of Fig. \ref{fig:SSBs}, so that the remaining resources within the 5 ms burst set can be used for data or other control channels.  

The speed of the tracking then depends on the UE receiver
capabilities. We consider two possible UE 
beamforming architectures: analog or fully digital.  These two transceiver choices
were also evaluated in the context of initial access in
\cite{barati2015directional,barati2016initial}.
The front-end architecture choice remains a major
outstanding design
issue in mmWave systems
as there are significant implications for power and area.

From the perspective of search, 
the key limitation of analog beamforming is that
it can only ``look'' in one direction at a time.
Since we assume in this study that the UE
must search 12 possible directions,
a full cycle takes $12 \times 20=240$ ms.
The gNB and UE update their beamforming vector pair whenever a stronger received power is detected.
In contrast, in digital beamforming, the UE can ``look'' in all
directions simultaneously and thus is able to measure all directions
every 20~ms. 

\section{Simulation Results and Discussion}

\begin{figure}[!t]
\input{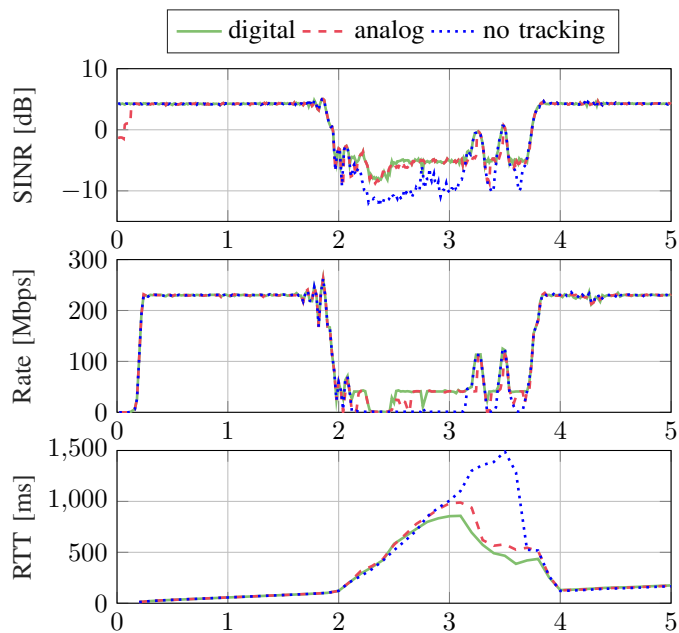}
\caption{Simulated TCP traffic over a measurement-based channel with different beam tracking schemes}
\label{sim}
\end{figure}

To evaluate the end-to-end performance with blockage,
we simulate sending TCP traffic to the UE whose mmWave channel and beam search procedure 
is described above. 
We use a 400~MHz bandwidth and full buffer traffic.  
The RLC buffer size is configured to be 5 MB and the core network round-trip delay is set to 10 ms. The results are shown in Fig.~\ref{sim} for three different beam tracking schemes:
\begin{enumerate}
\item{\bf Digital transceiver at the UE:} the UE performs digital beamforming, meaning that it will capture the best beam pair at every SS burst set.
\item{\bf Analog transceiver at the UE:} the UE performs analog beamforming, meaning that it will take a number of SS burst sets before discovering the best beam pair.
\item{\bf No tracking:} the UE detects the best beam pair in the initial access stage and uses this pair across the entire simulation.
\end{enumerate}


In the SINR plot, we observe 
three blockage events (at 2s-3s, 3.2s-3.3s and 3.4s-3.6s). The no-tracking line shows that 
the SINR drops around 14~dB when the link breaks. With digital beamforming, the SINR drops by approximately 
9~dB thanks to the ability to track the reflected path. The analog case can also utilize the reflected path, but has a small additional delay in finding the path
compared to the digital case. 

The rate is almost the same when the primary path is available. However, during blockage events, the digital case achieves a higher and smoother rate, which is consistent with the SINR plot. The analog case is slightly worse, but still produces a much higher rate compared to the no-tracking case.

The RLC buffer size is large enough to prevent overflow, but inevitably there was increased latency during blockage events. For all three beamtracking schemes the Round Trip Time (RTT) has the same behavior when there is no blockage. The trend differs between the three methods when the line-of-sight path is blocked. As expected, fully digital beamforming provides slightly lower latency than analog beamforming, due to a more timely reaction to the blockage event. In contrast, the no-tracking case barely transmitted any data during the blockage event and therefore accumulated more packets in the buffer and experienced the highest latency.


\section*{Conclusions}

In this paper, we have presented a detailed measurement-based study of end-to-end performance over mmWave links during blockage.
The study incorporates real spatial-temporal channel measurements into a thorough simulation.
We see that blockage can indeed dramatically impair the end-to-end performance, due to the beam search delays
and interactions with TCP.  Fully digital beamforming
can somewhat reduce the effects of blockage
suggesting that power efficient methods for realizing such front-end architectures may be worth investigating.

\bibliographystyle{IEEEtran}
\bibliography{bibl}

\end{document}